# Contributions of magnetic structure and nitrogen to perpendicular magnetocrystalline anisotropy in antiperovskite $\varepsilon$-Mn$_4$N

Shinji Isogami*, Keisuke Masuda**, and Yoshio Miura

National Institute for Materials Science, Tsukuba 305-0047, Japan

*E-mail : isogami.shinji@nims.go.jp

**E-mail : masuda.keisuke@nime.go.jp

To study how nitrogen contributes to perpendicular magnetocrystalline anisotropy (PMA) in the ferrimagnetic antiperovskite Mn$_4$N, we examined both the fabrication of epitaxial Mn$_4$N films with various nitrogen contents and first-principles density-functional calculations. Saturation magnetization ($M_s$) peaks of 110 mT and uniaxial PMA energy densities ($K_u$) of 0.1 MJ/m$^3$ were obtained for a N$_2$ gas flow ratio ($Q$) of ~10% during sputtering deposition, suggesting nearly single-phase crystalline $\varepsilon$-Mn$_4$N. Segregation of $\alpha$-Mn and nitrogen-deficient Mn$_4$N grains were observed for $Q \approx 6\%$, which were responsible for a decrease in the $M_s$ and $K_u$. The first-principles calculations revealed that the magnetic structure of Mn$_4$N showing PMA was "type-B" having a collinear structure, whose magnetic moments couple parallel within the $c$-plane and alternating along the $c$-direction. In addition, the $K_u$ calculated using Mn$_{32}$N$_x$ supercells showed a strong dependence on nitrogen deficiency, in qualitative agreement with the experimental results. The second-order perturbation analysis of $K_u$ with respect to the spin-orbit **interaction revealed that not only spin-conserving but also spin-flip processes contribute significantly to the PMA in Mn₄N. We also found that both contributions decreased with increasing nitrogen deficiency, resulting in the reduction of $K_u$. It was noted that the decrease in the spin-flip contribution occurred at the Mn atoms in face-centered sites. This is one of the specific PMA characteristics we found for antiperovskite-type Mn₄N.**

## I. INTRODUCTION

Perpendicular magnetic anisotropy (PMA) in magnetoresistive devices such as magnetic tunnel junctions[1] has attracted significant attention in view of their potential application in spin-transfer torque (STT) magnetic random access memories.[2] Owing to their PMA with small saturation magnetization ($M_s$), Mn-based alloys with PMA such as $D0_{22}$-Mn₃Ga,[3-6] $D0_{22}$-Mn₃Ge,[7,8] and $L1_0$-MnAl[9-12] are regarded as satisfying the requirement for small critical STT-switching current density, $J_c \propto \alpha M_s t H_k$ (where $\alpha$, $t$, and $H_k$ denote the damping constant, ferromagnetic layer thickness, and anisotropy field, respectively), which is proportional to the PMA energy density ($K_u$). For example, the $K_u$ and $M_s$ values for sputter-deposited $D0_{22}$-MnGa films are reported to be ~1 MJ/m³ and 310 mT,[3] which are ~15% and ~20% of the values obtained for bulk $L1_0$-FePt,[13] respectively. First-principles density-functional calculations also predict that the spin-flip scattering process due to spin-orbit interaction is the key for PMA in the $D0_{22}$-Mn₃Ga.[14]

Mn₄N with the $\varepsilon$ phase has also long been known as a Mn-based PMA ferrimagnetic material with an antiperovskite structure described by the formula $ANB_3$, where A and B correspond to Mn(I) at corner sites and Mn(II) at face-centered sites, respectively (Fig. 1).[15] A neutron diffraction study identified a ferrimagnetic spin order with two distinct magnetic moments, 3.5 $\mu_B$ for Mn(I) and $-0.8$ $\mu_B$ for Mn(II), at 300 K.[16] The $Mn_4N$ films recently fabricated by sputtering and molecular-beam epitaxy (MBE) exhibited $M_s \approx 110$ mT and $K_u \approx 0.1$ $MJ/m^3$.[17-20] Because its magnetic properties resemble those of the other Mn-based alloys mentioned above, recent research has focused on $Mn_4N$ in view of its potential for spintronics applications.[21-24] According to previous studies, the high $K_u$ of $Mn_4N$ films can be attributed to tetragonal distortion with $c/a \approx 0.99$, where $a$ and $c$ correspond to the in-plane and out-of-plane lattice constants, respectively.[19] Furthermore, the theoretical calculation of the crystal formation energy reveals that the free-standing $Mn_4N$ antiperovskite cell favors a collinear magnetic structure when $c/a$ approaches 0.99.[25]

The correlation between $c/a < 1$ and PMA is encountered in other Mn-based PMA alloys;[3] however, no theoretical study has yet been conducted on the magnetic structures of $Mn_4N$ that lead to $c/a < 1$ and to the resultant PMA. Furthermore, the magnetic moment of Mn atoms shows a strong site dependence originating from the interaction between Mn and nitrogen atoms.[15,16] However, the specific roles of nitrogen atoms in PMA are still under debate. Thus, the purpose of this study is to clarify, firstly, the magnetic structure of $Mn_4N$ showing high $K_u$, and secondly, the specific contribution of nitrogen to PMA.

In the present study, we prepared Mn$_4$N films and measured the variation of $K_u$ as a function of nitrogen content. Detailed structural analysis was performed to investigate the crystal grain growth and the chemical ordering of nitrogen. We also carried out first-principles calculations of $K_u$ for two possible collinear magnetic structures for Mn$_4$N. The $K_u$ variations for the Mn$_{32}$N$_x$ supercell were analyzed as a function of $x$ in order to extract the contribution of nitrogen to PMA.

## II. EXPERIMENTAL AND COMPUTATIONAL PROCEDURES

In the sample preparation and characterization, 30-nm-thick Mn$_4$N films were grown on single-crystal MgO(001) substrates via the reactive nitridation sputtering technique at a substrate temperature of 450°C. To change the amount of nitrogen in the Mn$_4$N films, the Ar and N$_2$ gas flows were controlled by mass flowmeters, and the ratio of N$_2$ ($S_{N2}$) to Ar ($S_{Ar}$) is defined as $Q = \frac{S_{N2}}{S_{Ar}+S_{N2}}$. Structural analysis of the (001)-oriented Mn$_4$N crystal layers was performed by X-ray diffraction (XRD) with Cu$K_\alpha$ radiation. Magnetic properties were measured using a superconducting quantum interference device vibrating sample magnetometer (SQUID-VSM) at room temperature. Magnetic anisotropy was measured via torque magnetometry based on the angle-dependent anomalous Hall effect (AHE).[26]

In the theoretical calculation of $K_u$, we carried out first-principles density-functional calculations with the aid of the Vienna *ab initio* simulation program (VASP),[27,28] where lattice constants estimated from the in-plane and out-of-plane XRD profiles were used. Prior to the estimation of $K_u$, we first calculated the formation energy of Mn$_4$N ($E_{total}$) assuming two types of collinear ferrimagnetic structures. While the non-collinear ferrimagnetic structure of cubic Mn$_4$N has been examined,[29,30] we did not address this in the present study, because the collinear magnetic structure is essential for PMA in the case of our (001)-oriented Mn$_4$N with distortion along [001]. We next calculated the value of $K_u$ for the stable magnetic structure using the force theorem,[31,32] $K_u = (E_{[100]} - E_{[001]})/V$, where $E_{[100]}$ ($E_{[001]}$) is the sum of the eigenenergies of the unit cell with the magnetization along the [100] ([001]) direction and $V$ is the volume of the unit cell. The force theorem approximates the difference in the $E_{total}$ between different magnetization directions as that in the sum of eigenenergies, which is known to be reasonable in most systems with magnetocrystalline anisotropy.[31,32] Actually, we recently confirmed the reliability of the force theorem for the interfacial magnetocrystalline anisotropy in magnetic tunnel junctions.[33] Using the same method, we also estimated the values of $K_u$ in the supercells Mn$_{32}$N$_x$ ($x = 1 \sim 13$), based on which we can discuss the role of nitrogen in Mn$_4$N theoretically. To understand the results of these calculations, we further carried out second-order perturbation calculations of $K_u$ for the supercells with respect to the spin-orbit interaction.[34-36] Further details of our calculations are provided elsewhere.[37]

## III. RESULTS AND DISCUSSION

**A. Magnetic properties**

Figure 2(a) shows the magnetization curves for $Q = 9\%$, where the magnetic field ($H$) was swept along the [100] and [001] directions. Comparing the two curves, the magnetic easy axis was found to point along the [001] direction; that is, the studied $Mn_4N$ film showed sizeable PMA. Although several different magnetic moments generally coexist in the unit cell of ferromagnetic $Mn_4N$, the PMA is defined as the energy difference of the net magnetic moment pointing in between easy and hard axis, which is consistent with the case of ferromagnets. The $\mu_0 M_s$ was measured to be 110 mT, which was comparable to that of $\varepsilon$-$Mn_4N$ films fabricated by sputtering and MBE.[19,20] Figure 2(b) summarizes the $\mu_0 M_s$ as a function of $Q$. The peak value appeared at $Q = 9\%$, whereas it decreased at both lower and higher percentages. The results were in good agreement with a previous report;[20] therefore, possible reasons for the degradation of $\mu_0 M_s$ against $Q$ can be attributed to the coexistence of pure Mn and/or nano-crystals such as $Mn_3N_2$ with stoichiometric $Mn_4N$ as impurities. Figure 2(c) shows a representative saturated magnetic torque curve based on measurement of the AHE.[26] By fitting the curve with $T = -(K_{u1} + K_{u2})\sin 2\theta_M + (K_{u2}/2)\sin 4\theta_M$, the first- and second-order uniaxial magnetic anisotropy constants $K_{u1}$ and $K_{u2}$ were obtained, where $\theta_M$ denotes the magnetization angle with respect to the film plane normal, given by $\theta_M = \arccos[R_{AHE}(H) / R_{AHE}]$. The resultant $K_u$ values were found to be dominated by $K_{u1}$ (the $K_{u2}$ contribution can be neglected), and the peak value was comparable to that of $Mn_4N$ fabricated using MBE[19] and pulsed laser deposition.[38] Note that the self-demagnetization energy density ($\mu_0 M_s^2/2$) was estimated to be 4.8 kJ/m$^3$ using $\mu_0 M_s = 110$ mT, which was two orders of magnitude smaller compared with the $K_u$ of 0.1 MJ/m$^3$. These results show that the experimentally obtained $K_u$ for the Mn$_4$N film can be dominated by magnetocrystalline anisotropy. As with $\mu_0 M_s$, the $Q$ at which the peak $K_u$ appeared was 9%, so we infer that the optimum $Q$ giving the highest PMA is near 9%.

**B. Structural characterization**

To consider the correlation between $K_u$ and the crystal structures of our Mn$_4$N films, structural analysis was conducted via XRD. Figure 3(a) shows the out-of-plane XRD profiles for the substrate/Mn$_4$N(30 nm) samples. For $Q = 10\%$, the diffraction peaks at $2\theta/\omega \approx 22.9°$ and $46.8°$ correspond to the (001) superlattice and (002) fundamental peaks of Mn$_4$N crystals, respectively. These peaks were still observed for $Q = 6\%$; however, additional peaks emerged at $2\theta/\omega \approx 40.4°$. Transmission electron microscopy revealed that the additional peaks originated from two impurity phases, consisting of $\alpha$-Mn crystal grains and a Mn-O crystal layer formed at the top surface of the Mn$_4$N layer due to natural oxidation. In addition, the chemical ordering of the nitrogen atoms in the Mn$_4$N grains decreased with decreasing $Q$ (see Fig. A1 in Appendix). Figure 3(b) shows the in-plane XRD profiles with the incident angle $\omega = 0.4°$, where the X-ray scattering vector is pointing along the MgO[200] direction. The diffraction peaks from Mn$_4$N(100) and Mn$_4$N(200) were observed as in Fig. 3(a). The broad peak at $2\theta_\chi/\phi \approx 41°$ cannot be attributed to the $\alpha$-Mn crystal grains but instead to the naturally oxidized Mn-O top layer, because in-plane XRD measurement is generally sensitive to the film surface. These results led us to conclude that the decrease in $K_u$ for lower $Q$ can be attributed not only to the impurity phase $\alpha$-Mn but also to the increase in nitrogen deficiency compared to stoichiometric Mn$_4$N crystals. Although the mechanisms involved are complex, the point we address here is not the influence of the impurity $\alpha$-Mn grains but the difference between stoichiometric and nitrogen-deficient Mn$_4$N.

**C. Calculation of magnetic structure**

We will now present our first-principles calculation results. First, we determined the possible magnetic structure of the equilibrium state of Mn$_4$N. Neutron diffraction experiments revealed two types of magnetic structures with ferrimagnetic order,[16] type-A: (positive, negative) = (Mn(I), Mn(II)); and type-B: (positive, negative) = (Mn(II)X/Y, Mn(I) and Mn(II)Z). Therefore, calculation was carried out for the two types. Figure 4(a) plots the $E_{total}$ of the Mn$_4$N unit cell with different magnetic structures, namely, type-A and type-B. The insets show schematic illustration of each magnetic structure. $E_{total}$ of the type-B structure was clearly smaller than that of type-A for all $c/a$. The $c/a$ that gave the smallest $E_{total}$ was 0.998 for type-A and 0.976 for type-B, and the difference reached ~0.2 eV/cell. These results were in good agreement with a previous report.[25] Here, we will present the possible explanation for the stable type-B magnetic structure. For simplicity, cubic structure is assumed for both the type-A and type-B. Considering the exchange interaction between Mn atoms, we consider the following Heisenberg Hamiltonian,

$$H = J_1 \sum_{\langle i,j \rangle} \mathbf{S}_i \cdot \mathbf{S}_j + J_2 \sum_{\langle\langle i,j \rangle\rangle} \mathbf{S}_i \cdot \mathbf{S}_j \quad , \qquad (1)$$

where $J_1$ and $J_2$ represent the 1st nearest Mn-Mn and the 2nd nearest Mn-Mn or Mn-N-Mn exchange interaction constants, respectively. Note that the $J_n > 0$ ($J_n < 0$) corresponds to the antiferromagneitc (ferromagnetc) coupling. The location of the magnetic moment is denoted by $i$, and $<i, j>$ and $<< i, j >>$ indicate sums over 1st and 2nd nearest neighbor pairs of spins. The total classical energy of all $\mathbf{S}_i$ presented at atomic coordinates is given by,

$$E = J_1 S^2 \sum_{\langle i,j \rangle} \cos\theta_{ij} + J_2 S^2 \sum_{\langle\langle i,j \rangle\rangle} \cos\theta_{ij} \quad , \qquad (2)$$

where $\theta_{ij}$ represents the relative angle between two spins, $S_i$ and $S_j$. The $E$ of the type-A and the type-B were estimated with respect to the $\mathbf{S}_i$ of Mn at four atomic coordinates: $(a,b,c) = (0,0,0)$, $(1/2,1/2,0)$, $(1/2,0,1/2)$, and $(0,1/2,1/2)$. In the case of type-A, there are four bonds with $J_1$ and $J_2$ for each $xy$, $yz$, and $zx$ plane with multiplicity of two, therefore, the resultant energy of the $\mathbf{S}_i$ at $(0,0,0)$ ($E_{000}$) is given by,

$$E_{000} = S^2\big(J_1(4\cos\pi + 4\cos\pi + 4\cos\pi) + J_2(4\cos 0 + 4\cos 0 + 4\cos 0)\big) \times \frac{1}{2}$$

$$= S^2(-6J_1 + 6J_2). \qquad (3)$$

Taking into account the site equivalency, the other energy can be estimated similarly to the $E_{000}$ as,

$$E_{1/2\,1/2\,0} = E_{1/2\,0\,1/2} = E_{0\,1/2\,1/2} = S^2(2J_1 + 6J_2). \qquad (4)$$

Then, we obtain $E_{\text{type-A}} = 24J_2S^2$ in four sites total, suggesting that only the $J_2$ terms exist in the energy of type-A, whereas the $J_1$ terms were canceled out because of the parallel (symmetric) spin configuration for the Mn(II)X/Y/Z in the octahedral coordination structure. In contrast to the type-A, the $E_{000}$ of type-B is given by,

$$E_{000} = S^2(-2J_1 + 6J_2). \quad (5)$$

The $E_{1/2\ 1/2\ 0}$, $E_{1/2\ 0\ 1/2}$, and $E_{0\ 1/2\ 1/2}$ are the same as $E_{000}$, then we obtain $E_{\text{type-B}} = -8J_1S^2 + 24J_2S^2$ in four sites total, suggesting that the total energy of type-B includes not only $J_2$ but also $J_1$ terms. This is caused by the presence of low symmetric octahedral coordination consisting of the Mn(II)Z and the Mn(II)X/Y with antiparallel spin configuration. Comparing the $E_{\text{type-A}}$ with $E_{\text{type-B}}$, we find the relationship: $E_{\text{type-A}} > E_{\text{type-B}}$ ($E_{\text{type-A}} < E_{\text{type-B}}$) in the case of $J_1 > 0$ ($J_1 < 0$), independent of $J_2$. Namely, the $E$ of type-A and type-B via spin density functional theory shown in Fig. 4(a) can be explained by the presence of preferential antiferromagnetic (AFM) spin order for the 1st nearest Mn-Mn bonds, which was more remarkable for the type-B than the type-A. Such the AFM preferential spin order might be caused by the direct exchange coupling mechanism, in an analogy with CrN systems discussed by Filippetti et al.[39] They find $J_1 = 9.5$ meV (AFM) by the Cr-Cr direct interactions and $J_2 = -4$ meV (FM) by the Cr-N-Cr interactions. Given the similarities of Cr and Mn ions, the Mn-Mn direct interaction with AFM spin order might be a key to give a stable magnetic structure of type-B. Although the data is not shown, we additionally examined the $E$ of fcc-Mn unit cell via the same methods for both the type-A and type-B magnetic structures, but removing the nitrogen atom from the Mn$_4$N unit cell. The type-B showed stable compared with the type-A, which was consistent with the case of Mn$_4$N. These results show the presence of Mn-Mn AFM direct interaction as the Cr-Cr bonds. The tetragonal distortion along *c*-axis could provide somewhat influences to the *E*; however, the effects is not so strong for the present *c/a* regime [Fig. 4(a)] that the energy difference between the type-A and B (~0.2 eV/cell) can be explained. We thus infer such the wide remarkable energy difference can be dominantly attributed to the preferential spin order with type-B.

**D. Calculation of magnetic moment**

We calculated the magnetic moment for each Mn atom in the Mn$_4$N unit cell with type-A and type-B (Table I). The magnetic moment of Mn(I) showed a large negative value, whereas a large positive value was exhibited for Mn(II)X/Y, resulting in the relatively small saturation magnetization (*M*$_s$) of 179.7 mT for the type-B. In contrast, the type-A resulted in 7.86 mT, suggesting a negligibly small value. Comparing *M*$_s$ = 110 mT measured for the 30-nm-thick Mn$_4$N films and 179.7mT obtained theoretically for the type-B structure, possible reasons for the discrepancy are: the imperfect degree of order of nitrogen (*S* = 0.79); and the formation of the initial growth layer with different magnetic structure from type-B and the top-surface natural oxidation layer consisting of Mn-O. The *M*$_s$ value tends to decrease by the imperfect degree of order of nitrogen, due to the transformation of magnetic structures and/or the nanocrystals of another phase with negligibly small *M*$_s$.[20] The calculated *M*$_s$ (179 mT) for $S = 1.0$ thus decreases down to 141 mT when assuming $S = 0.79$. The further discrepancy between 141 mT and the measured 110 mT can be attributed to somewhat dead layer formation such as an initial growth layer near the MgO substrate and an Mn-O top surface layer with a thickness of around 3 nm judging from the cross-sectional TEM observation (Fig. A1 in Appendix). Based on our previous study,[40] it is revealed that many dislocations are generally formed near an interface between an MgO substrate with a thickness of around 2 nm and a transition metal nitride such as $Fe_4N$, of which crystal structure is the same as $Mn_4N$. The dislocations would give rise to the transformation of magnetic structure from the stable type-B to the other one such as type-A and/or $\alpha$-Mn; therefore, the $M_s$ of the initial growth layer might be neglected. In terms of the Mn-O layer on top, negligible small $M_s$ can be expected because of its antiferromagnetism.[41] We can speculate that the initial growth layer and Mn-O top layer act as dead layers whose thickness is around 5 nm in total in the present samples. When we tried to consider these dead layers in the estimation of $M_s$, then resultant ~117 mT (consistent with the measured $M_s$ of 110 mT) was obtained. The estimation led us to conclude that the discrepancy between the 110 mT and 178 mT is distinct; however, which is explained by the imperfect $S$ as well as the ~5-nm-thick top/bottom dead layer formation.

The noticeable characteristics realized here are the magnetic moment for Mn(II)Z, which is close to the values of Mn(II)X/Y for the type-A, whereas significantly different for the type-B. This can be attributed to the near cubic structure for the type-A, whereas tetragonal distortion ($c/a \sim 0.976$) for the type-B in the Mn$_4$N unit cell. The electron number of Mn ion is generally responsible for the magnetic moment in the case of Mn$_4$N, we thus infer that the number of electron for Mn(II)Z decreases by the stronger hybridization with neighboring N atom in the case of type-B, which is resulted from short interatomic distance between the Mn(II)Z and N due to the tetragonal distortion.

**Table I** Calculated magnetic moment of Mn(I), Mn(II)X, Mn(II)Y, and Mn(II)Z in the Mn$_4$N unit cell with type-A and type-B magnetic structures.

|  | Mn(I) ($\mu_B$) | Mn(II)X/Y ($\mu_B$) | Mn(II)Z ($\mu_B$) | $M_s$ (mT) |
| --- | --- | --- | --- | --- |
| Type-A | 3.67 | –1.38/–1.47 | –1.58 | 7.86 |
| Type-B | –3.44 | 2.74 | –0.98 | 179.7 |

**E. Dependence of $K_u$ on $c/a$**

Figure 4(b) shows the $c/a$ dependence of the $K_u$ values for both the type-A and type-B structures calculated by using the Mn$_4$N unit cell. Here, we used $31 \times 31 \times 31$ $k$ points for the estimation of $K_u$ after confirming the convergence of $K_u$ against the number of $k$ points [Fig. 4(c)]. The $c/a$ values used for this calculation were based on the XRD analysis for the Mn$_4$N films with $Q$ = 5%, 10%, and 15% (solid symbols) (Fig. A2 in Appendix). The $E_{total}$ of the type-B structure was found to be smaller than that of type-A for the entire $c/a$ range, and the minimum $E_{total}$ was obtained at $c/a$ =

0.998 (type-A) and 0.976 (type-B) [Fig. 4(a)]. We also calculated $K_u$ at $c/a = 0.998$ for the type-A structure and at $c/a = 0.976$ for the type-B structure, where $E_{total}$ had the minimum values (open symbols). It is important to note that the $K_u$ of the type-B structure has positive values, indicating PMA, whereas that of the type-A structure has negative values, indicating in-plane magnetic anisotropy. Therefore, it can be concluded that the magnetic structure of Mn$_4$N films fabricated with $Q \approx 10\%$ and exhibiting high PMA is "type-B". In addition, the $K_u$ values were not strongly influenced by the $c/a$ ratio, so that the crystal distortion of the type-B structure cannot fully explain the dependence of $K_u$ on $Q$ observed in Fig. 2(d).

**F. Dependence of $K_u$ on the nitrogen deficiency**

Figure 5 shows the dependence of $K_u$ on the number of nitrogen atoms $x$ calculated using the Mn$_{32}$N$_x$ supercell, whose size was 2×2×2 that of the Mn$_4$N unit cell. Here, we used 15×15×15 $k$ points to estimate $K_u$, which is large enough because $K_u$ was confirmed to converge for $k$ points larger than 25×25×25 in the case of the Mn$_4$N unit cell (1×1×1) [Fig. 4(c)]. We set the $c/a$ ratio to 0.990, which is the experimental value that gave maximum $K_u$ [Fig. 2(d) and Fig. A2(b) in Appendix], and the type-B magnetic structure was employed since we confirmed lower $E_{total}$ for the type-B magnetic structure than for the type-A in the cases of Mn$_{32}$N$_8$ and Mn$_{32}$N$_1$. As a result, the $K_u$ value for $x \approx 8$ was calculated to be ~3.8 MJ/m$^3$, which dropped to ~2.6 MJ/m$^3$ with decreasing $x$. Note that the dependence of $K_u$ on $x$ was more prominent than that on the $c/a$ ratio (Fig. 4), suggesting that nitrogen deficiency dominates the mechanism for the degradation of $K_u$. The calculated $K_u$ value was nearly one order of magnitude higher than the experimental results. We will address the possible reasons as follows. Taking in to account that the degree of order ($S$) was estimated to be up to 0.79 [Fig. A2(c) in Appendix], which was close to that of the sputter-deposited Mn$_4$N films reported by another group,[20] and that degradation of $K_u$ due to shape anisotropy was negligible as mentioned in the description for Fig. 2(d), we speculate the reason for the discrepancy to be the imperfect chemical ordering of nitrogen involving the change in magnetic structure from type-B and/or the formation of dislocations in the initial growth layer.[40] The imperfect nitrogen order might partly form possible impurity phases such as nanocrystals of pure Mn and/or Mn$_4$N with different (type-A-like) magnetic structures from the type-B. These contaminations might decrease the entire PMA of Mn$_4$N layer, because the $K_u$ of contaminations are predicted to be negligible small, judging from Fig. 4(b). In terms of the initial growth layer of Mn$_4$N film deposited on the MgO substrate, many dislocations might be present due to large mismatch by ~ 9%. Such the Mn$_4$N initial layer with dislocations cannot qualitatively contribute to the PMA due to weak crystallinity, resulting in the degradation of $K_u$. Thus, the presence of nitrogen is presumably indispensable for obtaining a sufficiently large PMA in Mn-based antiperovskite nitrides such as Mn$_4$N. Nevertheless, the variation of $K_u$ with $x$ in the calculation agreed with the variation with $Q$ in the experiments [Fig. 2(d)]. Another reason to be considered is somewhat the reliability of employing the force theorem to estimate $K_u$ values of ferrimagnetic systems.

As we mentioned at the computational procedure, the force theorem has been widely known as a reasonable method in most systems with magnetocrystalline anisotropy,[31,32] even in that with an interfacial magnetocrystalline anisotropy.[33] However, one must consider that there might be some material systems that are beyond the reliability of existing force theorem. Although it is not clear, ferrimagnets might belong to such the systems owing to their completely different magnetic structures from usual FMs. Extensive study based on both experiments and calculation might be necessary to overcome these issues.

**G. Second-order perturbation analysis of PMA**

In order to determine how each Mn atom contributes to PMA due to the presence of nitrogen, we carried out a second-order perturbation analysis of $K_\mathrm{u}$ with respect to the spin-orbit interaction in the cases of $x = 1$ and 8, namely, $Mn_{32}N_1$ and $Mn_{32}N_8$. Figures 6(a-1) and 6(a-2) show the contributions to $K_\mathrm{u}$ from spin-conserving terms and spin-flip terms, respectively, at the Mn(I), Mn(II)X, Mn(II)Y, and Mn(II)Z sites of the $Mn_{32}N_8$ stoichiometric supercell. Here, the terms that can contribute to PMA are shown to be positive values, whereas those that contribute to in-plane magnetic anisotropy are negative; the values were averaged over eight atoms for each site. It was revealed that both the spin-conserving and spin-flip terms contribute to PMA: the spin-conserving term (the spin-flip term) has a relatively large value at the Mn(II)X/Y/Z (Mn(I)/Mn(II)X) atoms. To be specific, the spin-flip process from down- to up-spin state was dominant for the Mn(I) site, whereas that from up- to down-spin state dominated for the Mn(II)X site. Furthermore, the spin-flip process was suppressed only for Mn(II)Y and Mn(II)Z [Fig. 6(a-2)]. Note here that the site equivalence in our results between Mn(II)X and Mn(II)Y sites is lost, which is resulted from our calculation carried out under $M$ // [100] with the spin-orbit interaction: the spin moment in the Mn(II)X site points to the nitrogen atom when $M$ // [100] whereas that in the Mn(II)Y site does not. Therefore, the second-order perturbation results showed a strong site-dependent contribution to PMA owing to the presence of nitrogen atoms. This must be a characteristic of antiperovskite nitrides, such as Mn$_4$N, in which the nitrogen atom occupies the body-centered site. With nitrogen deficiency, in the case of Mn$_{32}$N$_1$, all the spin-conserving and the spin-flip terms drastically decreased [Figs. 6(b-1) and 6(b-2)]. These results show that the degradation of $K_u$ correlates not only with the spin-conserving terms but also with the spin-flip terms.

**H. Interpretation of PMA using atom-resolved DOS**

To examine the site-dependent spin-flip terms as well as their degradation due to nitrogen deficiency in more depth, we calculated the atom-resolved DOS for the Mn$_{32}$N$_8$ and Mn$_{32}$N$_1$ supercells [Figs. 7(a) and 7(b)]. Although the data is not shown here, we confirmed that the features of the DOS of the Mn$_{32}$N$_8$ supercell are almost the same as that of the Mn$_4$N unit cell. We found that the spin-flip contributions to $K_u$ shown in Fig. 6(a-2) agree with the processes expected from the DOSs as follows. In the case of Mn$_{32}$N$_8$ [Fig. 7(a)], the Mn(II)X/Y sites with positive magnetic moment have sharp peaks in the DOS around the Fermi level both in the occupied up-spin (–0.8 eV) and unoccupied down-spin (0.2 eV) states, which are composed of occupied d($xy$) and unoccupied d($zx$) orbitals. We confirmed that these orbitals provide a matrix element $<zx|Lx|xy>$ of spin-flip scattering from up-spin to down-spin channels, which positively contributes to the PMA of Mn$_4$N. In contrast to the Mn(II)X/Y sites, the Mn(I) site with negative magnetic moment has relatively large DOSs in the unoccupied up-spin states (0.5 eV), leading to a contribution to PMA by the spin-flip scattering from down to up spin channels. The DOS of the Mn(II)Z site is similar to that of the Mn(II)X/Y site, however, corresponding spin-flip scattering does not contribute to PMA but to in-plane magnetic anisotropy. This is consistent with the fact that a relatively large matrix element of $<xy|Lz|x^2-y^2>$ contributing to in-plane magnetic anisotropy presents in the Mn(II)Z site. To sum up, coexisting of atoms with opposite magnetic moments in ferrimagnets lead to a different spin-flip scattering, resulting in different contributions to PMA depending on each atom. This is a characteristic of PMA mechanism in ferrimagnetic Mn$_4$N revealed by the present second-order perturbation analysis. Note that the measured PMA by experiments corresponds to an effective one including all of the spin-flip scattering for each atom. Figure 7(b) shows the DOS of the Mn$_{32}$N$_1$ supercell. Compared with the case of Mn$_{32}$N$_8$, the clear peaks near the Fermi level were strongly smeared out [Fig. 7(b)], which is associated with the degradation of the spin-flip contribution due to the nitrogen deficiencies.

## IV. CONCLUSION

We have studied the role of nitrogen in the PMA of Mn$_4$N by both experiments and first-principles density-functional calculations. The results show that:

a) Not only the presence of impurity phases such as $\alpha$-Mn but also the decrease in the chemical ordering of nitrogen atoms, namely, nitrogen deficiency, is responsible for the degradation of $K_u$ as well as $M_s$.

b) Mn$_4$N showing PMA is associated with the "type-B" collinear structure with ferrimagnetic order.

c) Both the spin-conserving and spin-flip processes contribute to PMA. Furthermore, the contribution to PMA from each site was found to be clear owing to the presence of nitrogen: the spin-conserving process dominates the PMA component for the Mn(II)X, Mn(II)Y and Mn(II)Z sites, while spin-flip processes from down- to up-spin state for the Mn(I) site and from up- to down-spin state for the Mn(II)X site contribute to the PMA in Mn$_4$N.

d) Both the spin-conserving and the spin-flip processes decreased with decreasing $x$, which can explain the experimentally-observed degradation of $K_u$ in Mn$_4$N with N deficiency.

These are the characteristics we have identified for antiperovskite nitrides such as $\varepsilon$-Mn$_4$N.

**Appendix**

**A1. Transmission electron microscopy (TEM) observation for Mn$_4$N films**

In order to identify the coexistence of nano-crystals with different phases in the Mn$_4$N films, TEM observation was carried out. Figures A1(a1) and A1(a2) show the annular dark field (ADF) STEM image and the corresponding energy dispersive spectroscopy (EDS) mapping image for the Mn$_4$N films with $Q$ = 10%. The samples were covered with a thick Ni layer to prevent the sample from damage during the focused ion beam fabrication process. The Mn$_4$N layer shows a smooth surface, and EDS image indicates a nearly homogeneous nitrogen distribution within the imaged area. Note that the thin layer of a few nanometers adjacent to the Mn$_4$N surface could be identified as Mn-O crystals formed by natural oxidation, judging from the EDS mapping as well as the nanobeam diffraction (NBD) pattern [Figs. A1(a2) and A2(a3)]. The NBD of the Mn$_4$N layer suggested high chemical ordering of nitrogen [Fig. A1(a4)], because clear spots corresponding to a (001) superlattice appeared. We obtained similar NBD from several different points in the Mn$_4$N layer, so that $\varepsilon$-Mn$_4$N crystals grew homogeneously for $Q$ = 10%. In the case of $Q$ = 6%, bright and dark parts were observed in the STEM image [Fig. A1(b1)]. The contrast observed in the STEM image corresponds to nitrogen content. Figure A1(b2) shows EDS mapping, and we found that the bright part involved lower nitrogen content. To obtain the crystal structure, we observed the NBD as shown in Figs. A1(b3) and A1(b4). As a result, the bright part in the STEM image was identified as $\alpha$-Mn. The dark part of the STEM revealed Mn$_4$N grains; however, distinct nitrogen deficiency was present judging from the significantly weak or invisible NBD spots originating from the Mn$_4$N superlattice. These results show that pure $\alpha$-Mn and Mn$_4$N with remarkable nitrogen deficiency are formed as $Q$ decreases.

**A2. Lattice constants ($a$, $c$, and $c/a$) estimated using XRD profiles**

Figure A2(a) plots the lattice constants, $a$ and $c$, against $Q$ estimated using X-ray diffraction (XRD) profiles. The $a$ and $c$ indicate the lattice constants along in-plane and out-of-plane directions of the unit cell, as shown in Fig. 1 of the main text. In general, light elements such as nitrogen atoms act as interstitial impurities in nitrides. Therefore, both $a$ and $c$ monotonically increased with $Q$ as expected, which was similar to the results in previous report.[1] The resultant $c/a$ values decreased with increasing $Q$, which demonstrates that the Mn$_4$N unit cell undergoes tetragonal distortion [Fig. A2(b)]. We found that the value $c/a \approx 0.99$ was associated with the highest $K_u$ and $M_s$, as mentioned in the main text. Figure A2(c) summarizes the degree of order of nitrogen ($S$) estimated from the out-of-plane XRD results [Fig. 3(a) in the main text]. The detailed method of estimation is reported elsewhere[1,2]. The peak value of $S \approx 0.79$ appeared at $Q \approx 9\%$, which corresponds to the nitrogen content giving the highest $K_u$ and $M_s$.

**ACKNOWLEDGEMENTS**

The authors thank S. Mitani, Y. K. Takahashi, J. Uzuhashi, and T. Ohkubo for fulfilling discussions and detailed structural analysis. This work was supported by JSPS KANENHI Grant Nos. 19K04499, 18H03787, and 16H06332.

S. Isogami and K. Masuda contributed equally to this work.

**References**

[1]  S. Ikeda, K. Miura, H. Yamamoto, K. Mizunuma, H. D. Gan, M. Endo, S. Kanai, J. Hayakawa, F. Matsukura, and H. Ohno, Nat. Mater. **9**, 721 (2010).

[2]  J. C. Slonczewski, J. Magn. Magn. Mater. **159**, L1 (1996).

[3]  B. Balke, G. H. Fecher, J. Winterlik, and C. Felser, Appl. Phys. Lett. **90**, 152504 (2007).

[4]  F. Wu, S. Mizukami, D. Watanabe, H. Naganuma, M. Oogane, Y. Ando, and T. Miyazaki, Appl. Phys. Lett. **94**, 122503 (2009).

[5]  H. Kurt, K. Rode, M. Venkatesan, P. Stamenov, and J. M. D. Coey, Phys. Rev. B **83**, 020405(R) (2011).

[6]  K. Wang, E. Lu, J. W. Knepper, F. Yang, and A. R. Smith, Appl. Phys. Lett. **98**, 162507 (2011).

[7]  S. Mizukami, A. Sakuma, A. Sugihara, T. Kubota, Y. Kondo, H. Tsuchiura, and T. Miyazaki, Appl. Phys. Express **6**, 123002 (2013).

[8]  S. Wurmehl, H. C. Kandpal, G. H. Fecher, and C. Felser, J. Phys.: Condens. Matter. **18**, 6171 (2006).

[9]  M. Hosoda, M. Oogane, M. Kubota, T. Kubota, H. Saruyama, S. Iihama, H. Naganuma, and Y. Ando, J. Appl. Phys. **111**, 07A324 (2012).

[10]  T. Sands, J. P. Harbison, M. L. Leadbeater, S. J. Allen, Jr., G. W. Hull, R. Ramesh, and V. G. Keramidas, Appl. Phys. Lett. **57**, 2609 (1990).

[11]     W. Van Roy, J. De Boeck, H. Bender, C. Bruynseraede, A. Van Esch, and G. Borghs, J. Appl. Phys. **78**, 398 (1995).

[12]     T. Sato, T. Ohsuna, and Y. Kaneko, J. Appl. Phys. **120**, 243903 (2016).

[13]     T. Klemmer, D. Hoydick, H. Okumura, B. Zhang, and W. A. Soffa, Scr. Metall. Mater. **33**, 1793 (1995).

[14]     W. S. Yun, G. B. Cha, I. G. Kim, S. H. Rhim, and S. C. Hong, J. Phys.: Condens. Matter. **24**, 416003 (2012).

[15]     W. J. Takei, G. Shirane, and B. C. Frazer, Phys. Rev. **119**, 1893 (1960).

[16]     W. J. Takei, R. R. Heikes, and G. Shirane, Phys. Rev. **125**, 1893 (1962).

[17]     K. M. Ching, W. D. Chang, T. S. Chin, J. G. Duh, and H. C. Ku, J. Appl. Phys. **76**, 6582 (1994).

[18]     S. Nakagawa and M. Naoe, J. Appl. Phys. **75**, 6568 (1994).

[19]     Y. Yasutomi, K. Ito, T. Sanai, K. Toko, and T. Suemasu, J. Appl. Phys. **115**, 17A935 (2014).

[20]     K. Kabara, and M. Tsunoda, J. Appl. Phys. **117**, 17B512 (2015).

[21]     X. Shen, A. Chikamatsu, K. Shigematsu, Y. Hirose, T. Fukumura, and T. Hasegawa, Appl. Phys. Lett. **105**, 072410 (2014).

[22]     K. Kabara, M. Tsunoda, and S. Kokado, AIP Advances **7**, 056416 (2017).

[23]     S. Isogami, A. Anzai, T. Gushi, T. Komori, T. Suemasu, Jpn. J. Appl. Phys. **57**, 120305 (2018).

[24]     T. Komori, T. Gushi, A. Anzai, L. Vila, J. P. Attane, J. Vogel, S. Isogami, K. Toko, and T.

Suemasu, J. Appl. Phys. **125**, 213902 (2019).

[25]     K. Ito, Y. Yasutomi, K. Kabara, T. Gushi, S. Higashikozono, K. Toko, M. Tsunoda, and T. Suemasu, AIP Advances **6**, 056201 (2016).

[26]     T. Ono, N. Kikuchi, S. Okamoto, O. Kitakami, and T. Shimatsu, Appl. Phys. Express **11**, 033002 (2018).

[27]     G. Kresse, and J. Hafner, Phys. Rev. B **47**, 558(R) (1993).

[28]     G. Kresse, and J. Furthmiiller, Comput. Mater. Sci. **6**, 15 (1996).

[29]     M. Uhl, S. F. Matar, and P. Mohn, Phys. Rev. B **55**, 2995 (1997).

[30]     C. Li, Y. Yang, L. Lv, H. Huang, Z. Wang, S. Yang, J. Alloys Compd, **457**, 57 (2008).

[31]     G. H. O. Daalderop, P. J. Kelly, and M. F. H. Schuurmans, Phys. Rev. B **41**, 11919 (1990).

[32]     M. Weinert, R. E. Watson, and J. W. Davenport, Phys. Rev. B **32**, 2115 (1985).

[33]     K. Masuda, and Y. Miura, Phys. Rev. B **98**, 224421 (2018).

[34]     P. Bruno, Phys. Rev. B **39**, 865(R) (1989).

[35]     G. V. D. Laan, J. Phys.: Condens. Matter. **10**, 3239 (1998).

[36]     G. Autès, C. Barreteau, D. Spanjaard, and M. C. Desjonquères, J. Phys.: Condens. Matter. **18**, 6785 (2006).

[37]     Y. Miura, S. Ozaki, Y. Kuwahara, M. Tsujikawa, K. Abe, and M. Shirai, J. Phys.: Condens. Matter. **25**, 106005 (2013).

[38]   X. Shen, A. Chikamatsu, K. Shigematsu, Y. Hirose, T. Fukumura, and T. Hasegawa, Appl. Phys. Lett. **105**, 072410 (2014).

[39]   A. Filippetti, and N. A. Hill, Phys. Rev. Lett. **85**, 5166 (2000).

[40]   I. Suzuki, J. Uzuhashi, T. Ohkubo, and S. Isogami, Mater. Res. Express **6**, 106446 (2019).

[41]   T. Ahmad, K. V. Ramanujachary, S. E. Lofland, and A. K. Ganguli, J. Mater. Chem. **14**, 3406 (2004).

**Figure captions**

**Figure 1.** Schematic of atomic arrangements in antiperovskite $ANB_3$ type $\varepsilon$-$Mn_4N$ unit cell used in the first-principles density-functional calculations, where A and B correspond to Mn(I) and Mn(II), respectively. The Mn(II) atoms arranged in $x$, $y$ and $z$ directions are described as Mn(II)X, Mn(II)Y and Mn(II)Z. They are distinguished in our second-order perturbation analysis of perpendicular magnetocrystalline anisotropy with respect to the spin-orbit interaction.

**Figure 2.** (a) Magnetization curves measured along [001] (red) and [100] (blue) directions for the MgO substrate/$Mn_4N$ (30 nm) sample, with the $N_2$ gas flow ratio $Q = 9\%$. (b) Saturation magnetization ($\mu_0 M_s$) as a function of $Q$. (c) Saturated magnetic torque curve for the $Mn_4N$ sample with $Q = 10\%$. (d) Variations of magnetic anisotropy constant, $K_u$, $K_{u1}$, and $K_{u2}$.

**Figure 3.** (a, b) Out-of-plane (a) and in-plane (b) XRD profiles for MgO sub./$Mn_4N$ (30 nm) samples fabricated using the reactive nitride sputtering method with various $N_2$ gas flow ratios ($Q$).

**Figure 4.** (a) Calculation of total formation energy ($E_{total}$) of $Mn_4N$ unit cell with different spin structures, type-A and type-B, as a function of $c/a$. (b) $c/a$ dependence of $K_u$ by first-principles calculations for the $Mn_4N$ unit cell. The insets represent different magnetic structures of type-A and type-B. The $c/a$ ratios for solid symbols correspond to the values estimated from XRD profiles with $Q$ = 3%, 10%, and 15% [Fig. A2(b) in Appendix]. The $c/a$ ratios for open symbols correspond to the values that gave the minimum energy for the Mn$_4$N unit cell [Fig. 4(a)]. (c) The number of $k$ points dependence of the $K_u$ for the type-B with $c/a = 0.99$.

**Figure 5.** Calculated $K_u$ as a function of the number of nitrogen atoms ($x$) in the Mn$_{32}$N$_x$ supercell with the type-B magnetic structure and $c/a = 0.99$.

**Figure 6.** (a, b) Second-order perturbation terms for Mn$_{32}$N$_8$ (a) and Mn$_{32}$N$_1$ (b) with the type-B magnetic structure and $c/a = 0.99$. (a-1, b-1) The spin-conserving contributions from up- to up-spin state and from down- to down-spin state are indicated by green and orange bars, respectively. (a-2, b-2) The spin-flip contributions from up- to down-spin state and from down- to up-spin state are indicated by blue and red bars, respectively.

**Figure 7.** (a, b) Atom-resolved density of states (DOS) for Mn$_{32}$N$_8$ supercell (a) and Mn$_{32}$N$_1$ supercell (b) with the stable magnetic structure of type-B and $c/a = 0.99$. The red, green, and red lines correspond to the DOS of Mn(I), Mn(II)X/Y, and Mn(II)Z, respectively.

**Figure A1.** (a1) Cross-sectional annular dark-field scanning transmission microscopy (ADF-STEM) image and (a2) energy dispersive spectroscopy (EDS) mapping image for MgO sub./Mn$_4$N (30 nm) fabricated by reactive sputtering with nitrogen gas flow ratio ($Q$) = 10%. (a3, a4) Nanobeam diffraction (NBD) patterns showing Mn-O (a3) and Mn$_4$N (a4). (b1–b4) are the same as (a1-a4) but for $Q$ = 6%.

**Figure A2.** (a,b,c) N$_2$ flow ratio ($Q$) dependence of lattice constants, $a$ and $c$ (a), $c/a$ (b) degree of order of nitrogen ($S$) (c) for the sputter deposited 30-nm-thick Mn$_4$N films.

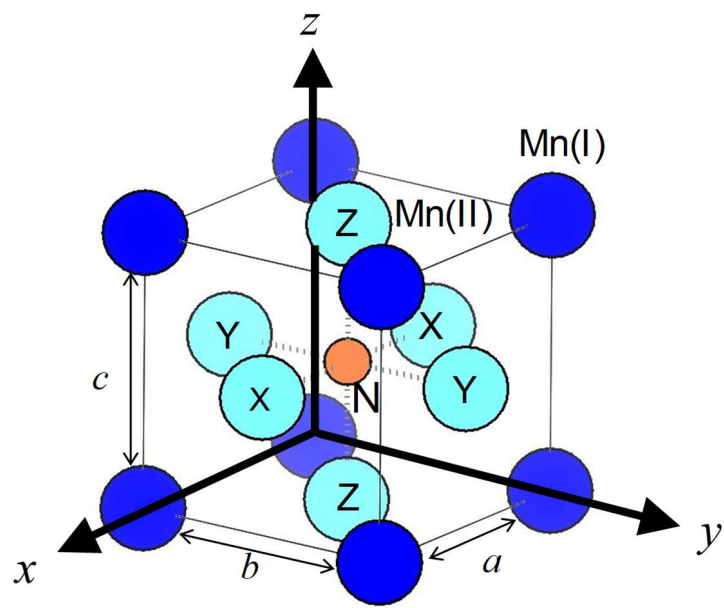

Fig. 1

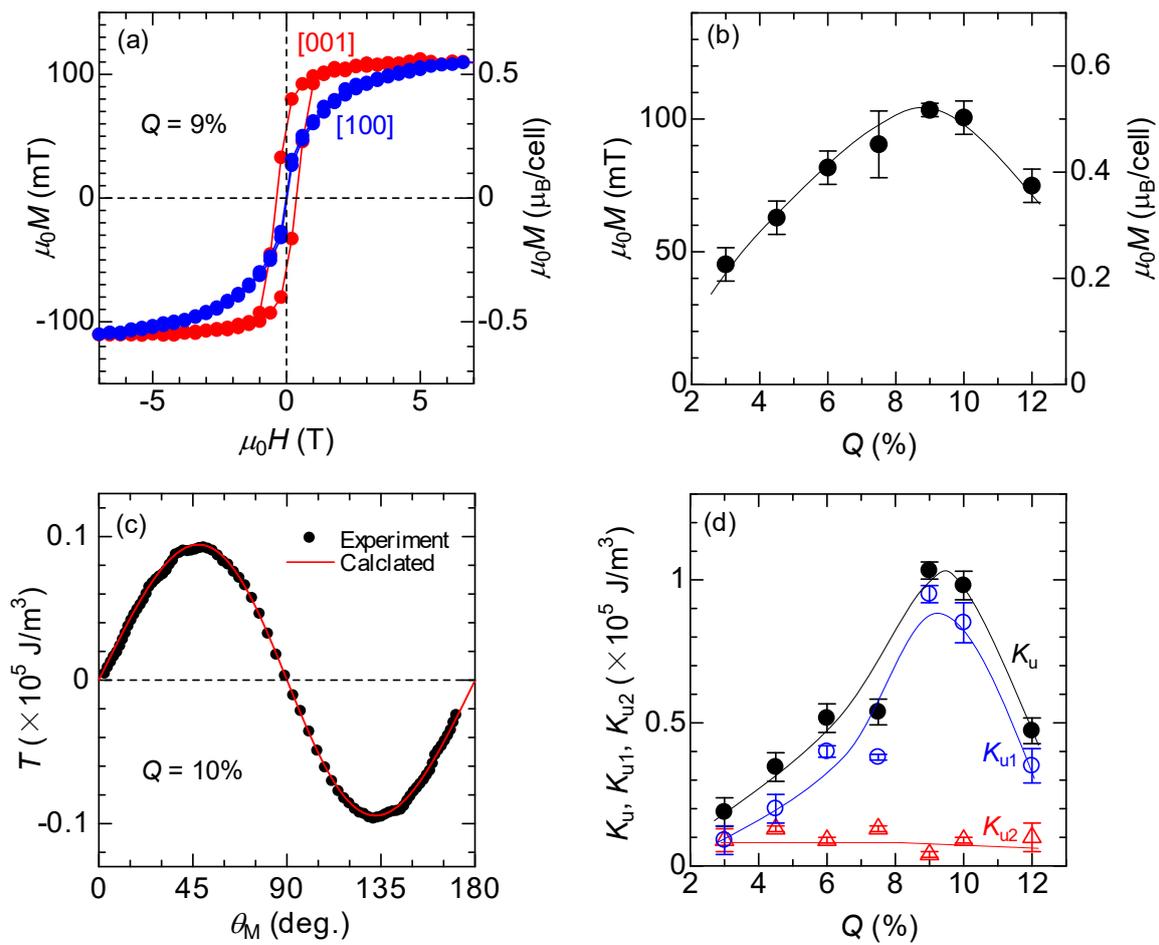

Fig. 2

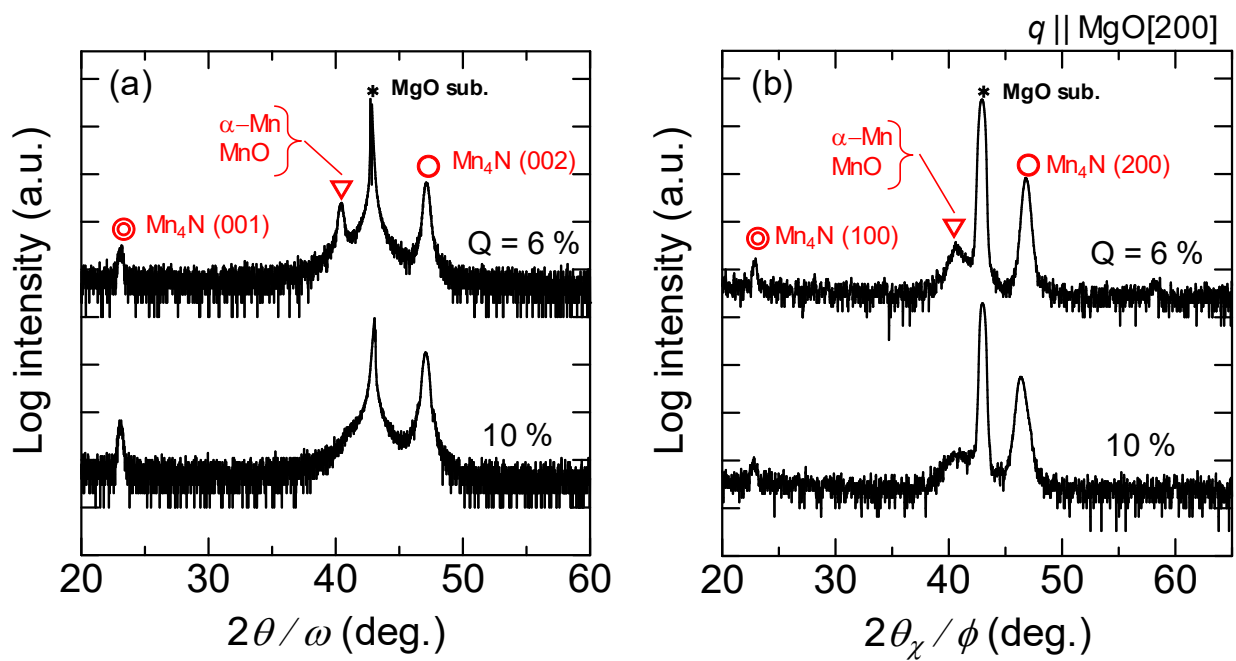

Fig. 3

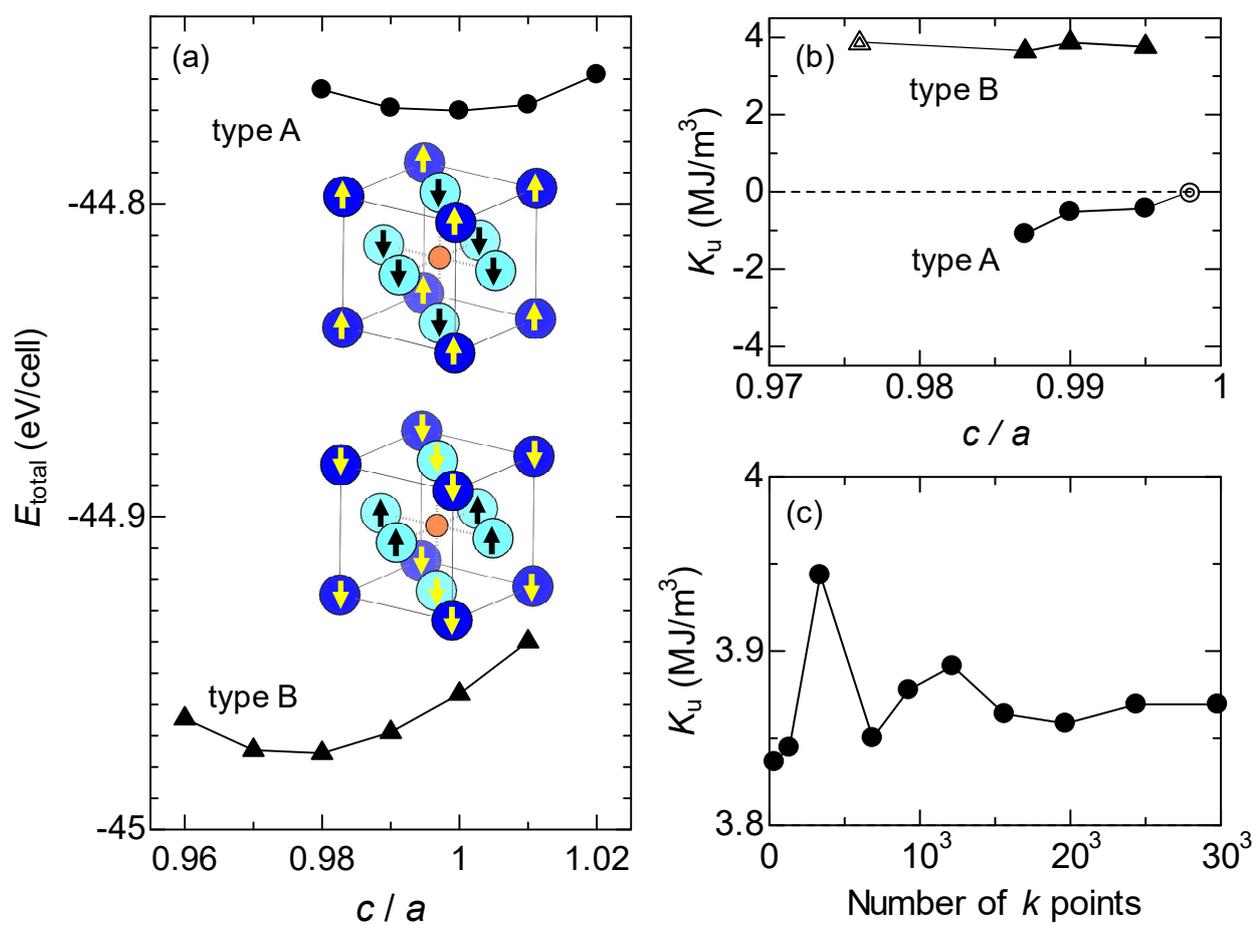

Fig. 4

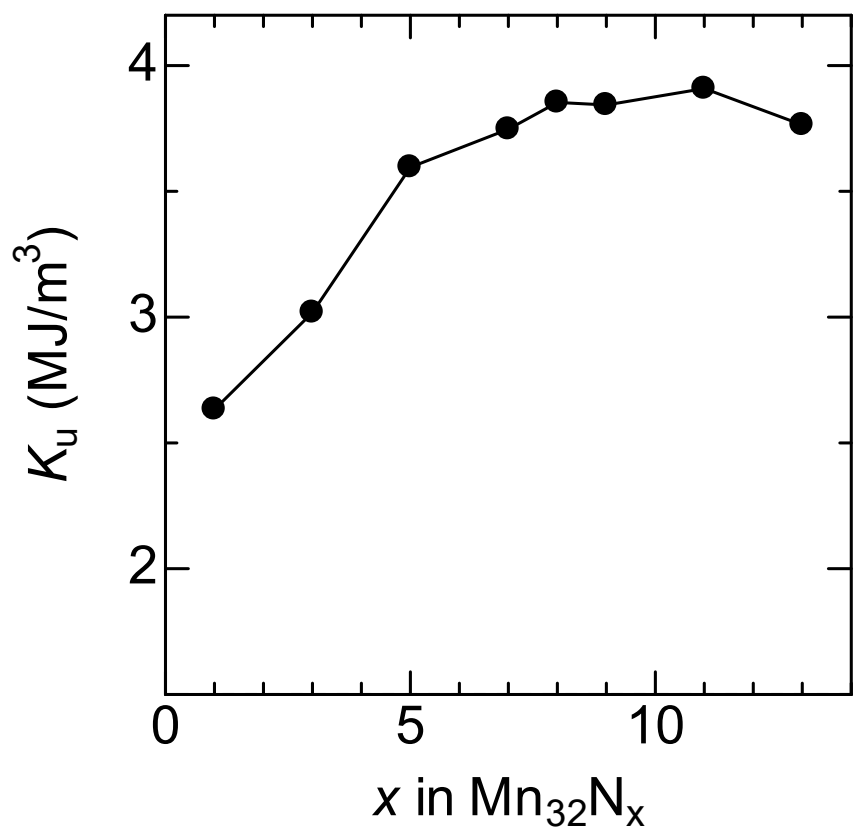

Fig. 5

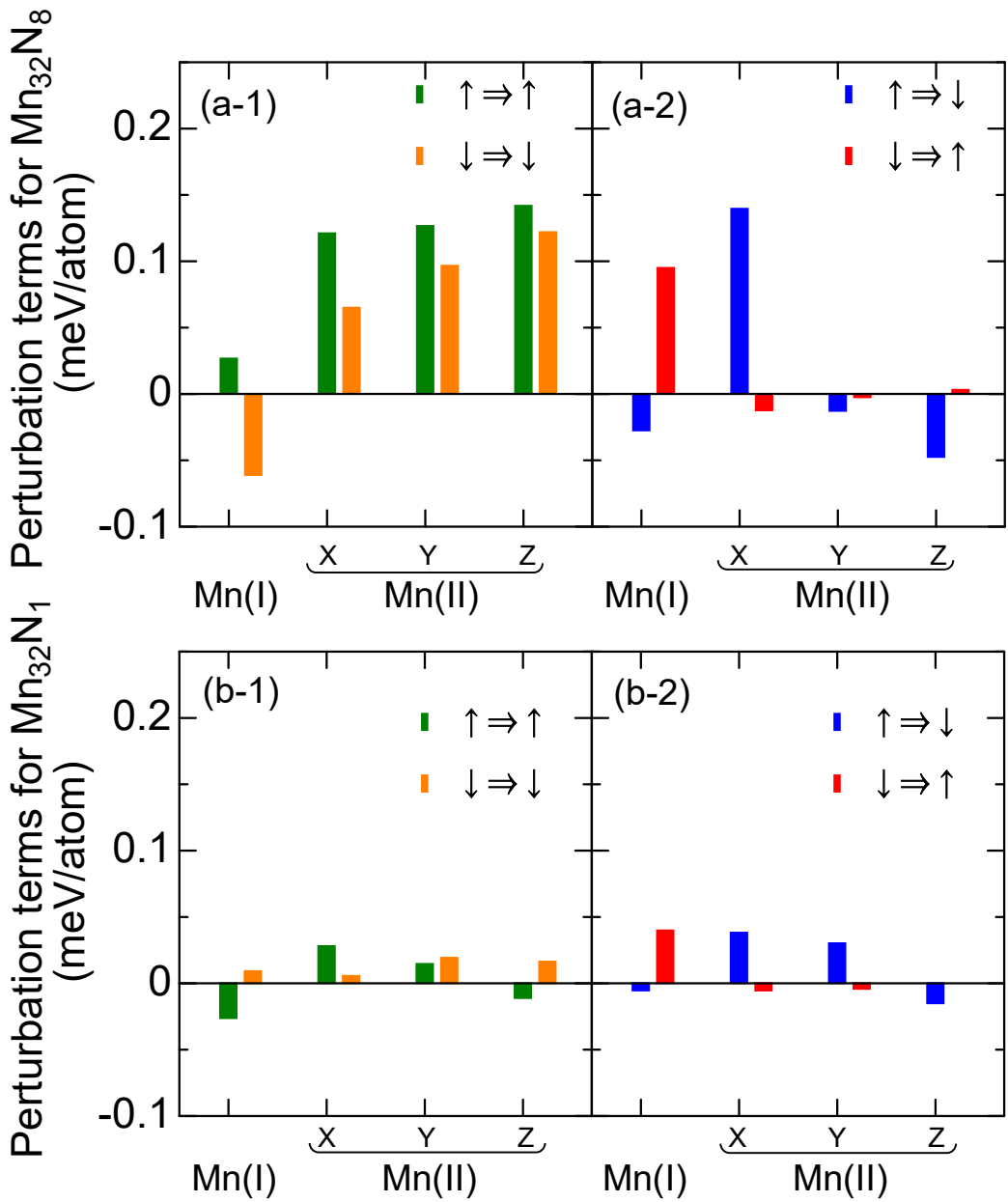

Fig. 6

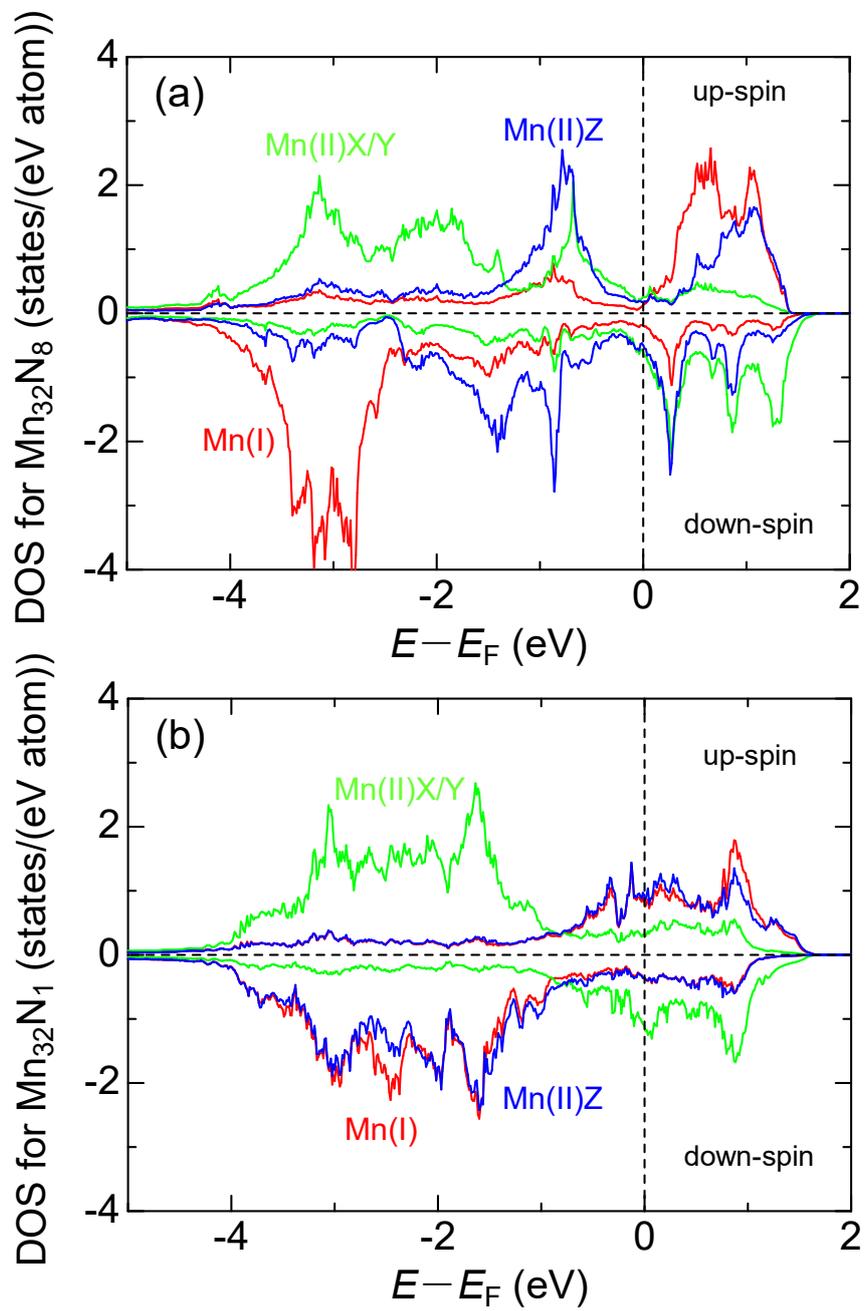

Fig. 7

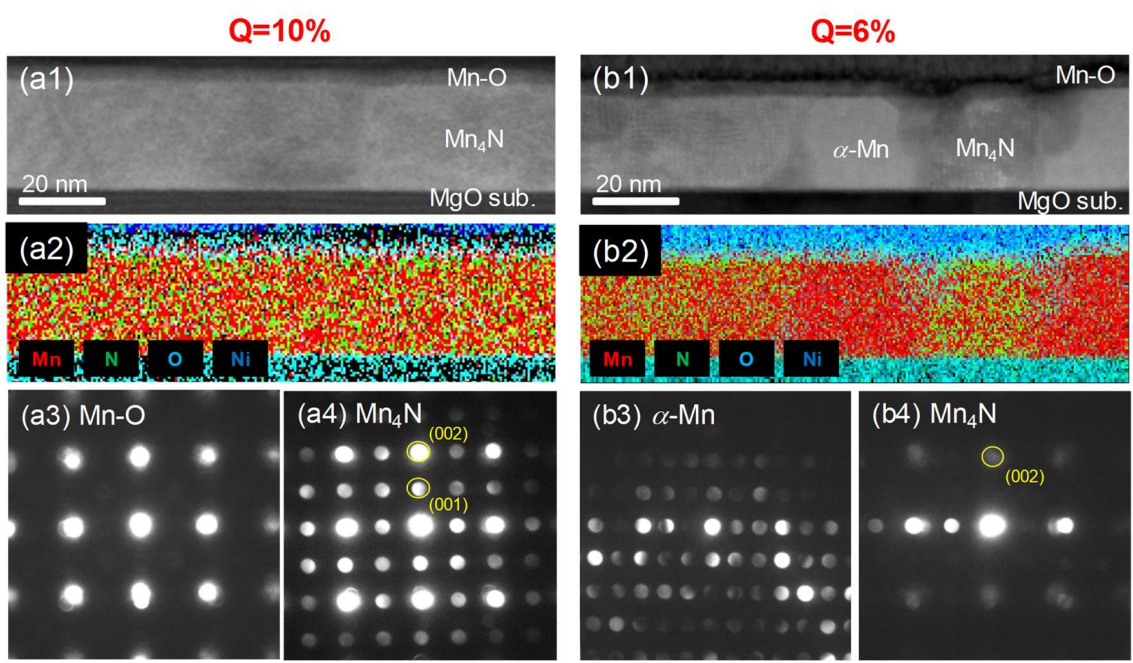

Fig. A1

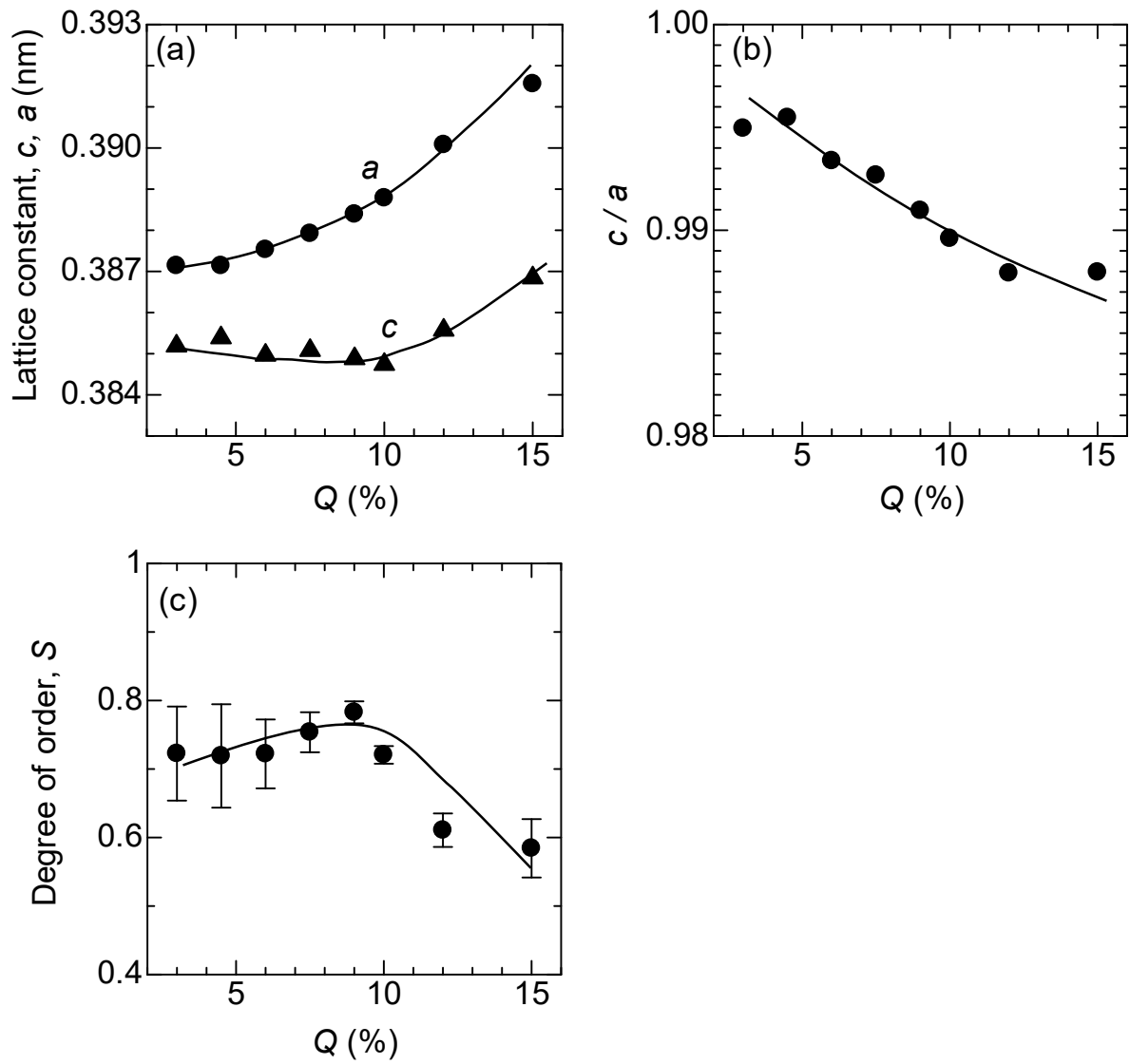

Fig. A2